\documentclass[aps,prb,amsmath,twocolumn,superscriptaddress,showpacs]{revtex4}

\usepackage{graphicx}
\usepackage{bm} 
\usepackage{url}

\DeclareMathOperator{\Img}{\mathrm{Im}}
\DeclareMathOperator{\Rea}{\mathrm{Re}}

\begin{document}
\author{Mikhail S. Kalenkov}
\affiliation{I.E. Tamm Department of Theoretical Physics, P.N. Lebedev Physical Institute, 119991 Moscow, Russia}
\author{Andrei D. Zaikin}
\affiliation{Institut f\"ur Nanotechnologie, Karlsruher Institut f\"ur Technologie
(KIT), 76021 Karlsruhe, Germany}
\affiliation{I.E. Tamm Department of Theoretical Physics, P.N. Lebedev Physical Institute, 119991 Moscow, Russia}
\title{Electron-hole imbalance and large thermoelectric effect in superconducting hybrids with spin-active interfaces}

\begin{abstract}
We argue that spin-sensitive quasiparticle scattering may generate electron-hole imbalance in superconducting structures,
such as, e.g., superconducting-normal hybrids with spin-active interfaces. We elucidate a transparent physical mechanism
for this effect demonstrating that scattering rates for electrons and holes at such interfaces differ from each other.
Explicitly evaluating the wave functions of electron-like and hole-like excitations in superconducting-normal bilayers
we derive a general expression for the thermoelectric current and show that -- in the presence of electron-hole imbalance --
this current can reach maximum values as high as the critical current of a superconductor.

\end{abstract}

\maketitle

\section{Introduction}

For several decades thermoelectric effect in superconductors was and remains one of the most intriguing topics of modern
condensed matter physics \cite{NL}. While theoretically this effect in ordinary superconductors is expected to be rather small
\cite{Galperin73}, a number of earlier experimental studies \cite{Zavaritskii74,Falco76,Harlingen80} indicated a much larger
result differing from theoretical predictions by several orders of magnitude. A similar conclusion was also reached
in a very recent experimental work \cite{Pe} although the reported discrepancy between theory and experiment appears to be somewhat
smaller in this case.

Which factors determine the magnitude of the thermoelectric effect in a metal? In the case of a normal metallic conductor,
simultaneous application of an electric field $\bm{E}$ and a temperature gradient $\nabla T$ yields an electric current
\begin{equation}
\bm{j}=\sigma_N \bm{E} + \alpha_N \nabla T,
\label{jN}
\end{equation}
where $\sigma_N$ is the standard Drude conductivity and $\alpha_N$ defines the thermoelectric coefficient of a normal metal.
Provided the temperature is sufficiently low and elastic electron scattering on non-magnetic impurities remains the dominant mechanism of its momentum relaxation, the thermoelectric coefficient $\alpha_N$ can be estimated by
means of the well known Mott formula
\begin{equation}
\alpha_N = \dfrac{2\pi^2}{9}eT
\dfrac{\partial}{\partial \mu} \left[N(\mu) \tau(\mu) v^2(\mu)\right]_{\mu=\varepsilon_F},
\label{mott}
\end{equation}
where $\varepsilon_F$ is the Fermi energy. This formula demonstrates that $\alpha_N$ may differ from zero only provided the product of the electron density of states $N$, its elastic scattering time $\tau$ and the square of its velocity $v$ substantially depends on energy in the vicinity of the Fermi surface. In generic metals, however, this dependence is usually pretty weak and, hence, the thermoelectric coefficient is typically small
$\alpha_N \sim (\sigma_N/e)(T/\varepsilon_F)$.

One can also demonstrate \cite{Galperin73} that the same small factor $T/\varepsilon_F \ll 1$ also controls the thermoelectric coefficient $\alpha_S$
in superconductors. In this case Eq. \eqref{jN} does not apply anymore, since no electric field can penetrate into a superconductor. Instead, a supercurrent
$\bm{j}_s$ can be induced by applying a temperature gradient to the system. In uniform superconductors this supercurrent is exactly compensated by the
thermoelectric current $\bm{j}_s =- \alpha_S \nabla T$, i.e. the net current
just vanishes in this case. In contrast, in non-uniform structures, such as, e.g., bimetallic rings, no such compensation is expected \cite{Ginzburg44,Ginzburg91} and, hence, such structures can be employed in order to experimentally investigate the thermoelectric effect in superconductors.

\begin{figure}
\centerline{ \includegraphics[width=80mm]{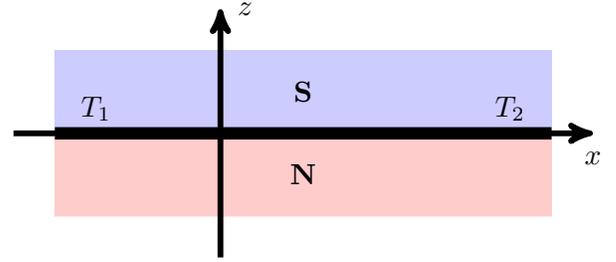} }
\caption{(Color online) SN bilayer with spin-active interface.}
\label{sfn-fig}
\end{figure}

Note that the above arguments explaining small values of the thermoelectric coefficient both in normal metals and superconductors apply only provided 
electron-hole asymmetry is weak in such systems. If, however, the symmetry between electrons and holes
is violated, one can expect a dramatic increase of thermoelectric currents.
Recently it was demonstrated that this is indeed the case, for instance, in conventional superconductors doped by magnetic impurities \cite{Kalenkov12},
in unconventional superconductors with quasibound Andreev states near non-magnetic impurities \cite{LF} or in
superconductor-ferromagnet hybrids with the density of states spin-split by the 
exchange or
Zeeman fields \cite{Machon,Ozaeta}. In this paper we will consider a 
different
structure -- a superconducting-normal (SN) bilayer (see Fig. 1) with a spin-active interface separating two metallic layers. We will demonstrate
that scattering rates for electrons and holes at such interface -- being strongly energy dependent at the scale of a superconducting energy gap $\Delta \ll \varepsilon_F$ -- may drastically differ from each other thereby generating strong electron-hole imbalance in the system. As a result, one can observe
a dramatic enhancement of the thermoelectric effect which may result in
huge thermoelectric currents reaching maximum values of order of the
critical (depairing) current of a superconductor.

\section{The model and basic formalism}
In order to proceed let us consider a 
metallic bilayer consisting of superconducting (S) and normal (N) slabs, as it 
is shown in Fig. \ref{sfn-fig}.  As we already pointed out, in what follows we 
will assume that these S- and N-metals are separated by a spin-active interface 
which can be produced, e.g., by an ultrathin layer of a ferromagnet. For the 
sake of simplicity here we will merely address the case of clean metals in 
which quasiparticles move ballistically and can scatter only at the SN 
interface. Finally, we will assume that the left and right ends of our bilayer 
are maintained at temperatures $T_1$ and $T_2$ respectively (see Fig. 1). 
Hence, quasiparticles entering our system from the left (right) side are 
described by the equilibrium (Fermi) distribution function with temperature 
$T_1$ ($T_2$).

The wave functions of quasiparticles propagating in our system obey the
well known Bogolyubov-de Gennes equations
\begin{equation}
\begin{pmatrix}
- (1/2m) \nabla^2 - \mu & \Delta \\
\Delta^* & (1/2m) \nabla^2 + \mu
\end{pmatrix}
\begin{pmatrix}
u \\ v
\end{pmatrix}
=\varepsilon
\begin{pmatrix}
u \\ v
\end{pmatrix},
\label{BdG}
\end{equation}
together with the normalization condition
\begin{equation}
\int(u^+_{\lambda} u_{\lambda'} + v^+_{\lambda} v_{\lambda'})  d \bm{r} =
\delta(\lambda- \lambda').
\label{norm}
\end{equation}
Here $u$, $v$ represent the two-component spinors, $\lambda$ is the quantum number distinguishing different solutions, 
$\mu$ is the chemical potential and $\Delta$ is the superconducting order parameter which has no spin structure 
(i.e. it is proportional to unity matrix in the spin space which can be achieved by empolying an appropriate basis of states)
and which will be chosen real in our subsequent analysis. The current density in the system is expressed in the standard form
\begin{multline}
\bm{j}(\bm{r})
=
\dfrac{e}{2m}\sum_{\varepsilon_{\lambda}>0}
\Rea
\Bigl[
u_{\lambda}^+(\bm{r}) \hat{\bm{p}} u_{\lambda}(\bm{r})
n_{\lambda}
-\\-
v_{\lambda}^+(\bm{r})\hat{\bm{p}} v_{\lambda}(\bm{r})
(1-n_{\lambda})
\Bigr],
\label{tok}
\end{multline}
were $\hat{\bm{p}}=-i\nabla$ is the momentum operator, and $n_{\lambda}$ is the
occupation number for the state $\lambda$. In our model $n_{\lambda}$ just coincides with the equilibrium Fermi distribution function
corresponding to temperatures $T_1$ and $T_2$ respectively for the right and left moving quasiparticles.

The solutions of Eq. \eqref{BdG} both in a normal metal and in a superconductor are expressed as a superposition of incoming and outgoing waves
\begin{gather}
\begin{pmatrix}
u_N \\ v_N
\end{pmatrix}=
\sum_{\pm}
\begin{pmatrix}
u_{N\pm} \\ v_{N\pm}
\end{pmatrix}
e^{\pm i p_z z}
e^{i\bm{p}_{\parallel}\bm{\rho}},
\label{uN}\\
\begin{pmatrix}
u_S \\ v_S
\end{pmatrix}=
\sum_{\pm}
\begin{pmatrix}
u_{S\pm} \\ v_{S\pm}
\end{pmatrix}
e^{\pm i p_z z}
e^{i\bm{p}_{\parallel}\bm{\rho}},
\label{uS}
\end{gather}
where we defined $\bm{\rho}=(x, y)$, the quasiparticle momentum components parallel ($\bm{p}_{\parallel}$) and normal ($p_z = \sqrt{2m\mu - \bm{p}_{\parallel}^2} >0 $) to the SN interface, and introduced the envelope functions $u_{N,S\pm}$, $v_{N,S\pm}$ varying at scales much
longer than the Fermi wavelength.

In order to account for quasiparticle scattering at the spin-active SN interface we introduce the scattering matrix and
match the Bogolubov amplitudes at the interface by means of the equation
\begin{equation}
\begin{pmatrix}
u_{S+} \\ u_{N-} \\  v_{S+} \\ v_{N-}
\end{pmatrix}
=
\begin{pmatrix}
\hat{\mathcal{S}}^+ & 0 \\
0 & \hat{\mathcal{S}}^-
\end{pmatrix}
\begin{pmatrix}
u_{S-} \\ u_{N+} \\  v_{S-} \\ v_{N+}
\end{pmatrix},
\label{bou}
\end{equation}
where $\hat{\mathcal{S}}^{\pm}$  represent the normal state electron and hole interface S-matrices
\begin{equation}
\hat{\mathcal{S}}^{\pm}
=
\begin{pmatrix}
\hat R_{\pm\sigma}^{1/2} e^{\pm i\hat\theta/2} &
i \hat D_{\pm\sigma}^{1/2} e^{\pm i\hat\theta /2} \\
i \hat D_{\pm\sigma}^{1/2} e^{\pm i\hat\theta /2} & \hat R_{\pm\sigma}^{1/2} e^{\pm i\hat\theta/2}
\end{pmatrix},
\end{equation}
with $\hat D_{\pm\sigma} = 1- \hat R_{\pm\sigma}$ and
\begin{gather}
\hat R_{\sigma}=
\begin{pmatrix}
R_{\uparrow} & 0 \\
0 & R_{\downarrow}
\end{pmatrix},
\quad
\hat R_{-\sigma}=
\begin{pmatrix}
R_{\downarrow} & 0 \\
0 & R_{\uparrow}
\end{pmatrix}.
\end{gather}
Here $R_{\uparrow}$ and $R_{\downarrow}$ denote the
electron reflection coefficients respectively for the spin-up and spin-down directions, $\hat\theta =\theta \hat \sigma_3$ is $2\times 2$ diagonal matrix
in the spin space which accounts for the scattering phase $\theta$ and $\hat \sigma_3$ is the Pauli matrix.

\begin{figure}
\centerline{\includegraphics[width=80mm]{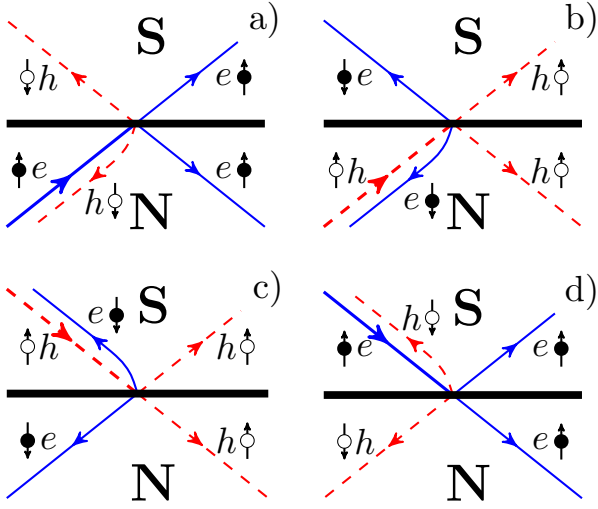}}
\caption{Four different electron and hole scattering processes in a
superconducting-normal bilayer.}
\end{figure}

\section{Electron-hole asymmetry} In order to construct a complete set of 
solutions of Eq. \eqref{BdG}
we will employ the standard scattering problem analysis and distinguish 16 different processes illustrated
in Fig. 2. Depending on whether incident electron-like or hole-like excitations come from the normal metal
or the superconductor one can classify all these processes into four groups labeled respectively
as (a), (b), (c) and (d) in Fig. 2. Consider, for instance, the four scattering processes of an electron-like excitation
arriving at the NS interface from the normal metal side. These four processes are depicted in Fig. 2a. Provided the energy
of this excitation $\varepsilon$ does not exceed $\Delta$, it cannot penetrate deep into the superconductor and gets reflected
back into the normal metal either in the form of an electron (specular reflection) or, alternatively, as a hole (Andreev reflection).
In the latter case, as usually, the charge conservation is assured by an extra Cooper pair going into the superconductor, implying
transferring the charge $2e$ across the NS interface. These processes are described by the wave functions \eqref{uN} if we choose
\begin{gather}
\begin{pmatrix}
u_{N+} \\ v_{N+}
\end{pmatrix}=
\begin{pmatrix}
1 \\ 0
\end{pmatrix}
e^{i\varepsilon z/v_z}+C_1
\begin{pmatrix}
0\\ 1
\end{pmatrix}
e^{-i\varepsilon z/v_z}
\label{uN1}
\end{gather}
(where $v_z=p_z/m >0$) and
\begin{gather}
\begin{pmatrix}
u_{N-} \\ v_{N-}
\end{pmatrix}=C_2
\begin{pmatrix}
1 \\ 0
\end{pmatrix}
e^{-i\varepsilon z/v_z}.
\label{uN2}
\end{gather}
Here the first and the second terms in the right-hand side of Eq. \eqref{uN1} account for the wave functions of respectively
an incident electron and a reflected hole while the wave function of a reflected electron is defined in Eq. \eqref{uN2}. Accordingly,
the reflection probabilities for both these processes are determined simply as $\mathcal{R}^{e-e}_{NS,\sigma}=|C_2|^2$ (normal reflection)
and $\mathcal{R}^{e-h}_{NS,\sigma}=|C_1|^2$ (Andreev reflection).

At electron energies $\varepsilon$ exceeding $\Delta$ in addition to the above two processes there also exist two extra ones: an electron
can penetrate into the superconductor from the normal metal both as an electron-like excitation and as a hole-like one, see Fig. 2a.
The latter process is again accompanied by creation of an extra Cooper pair in the superconductor, as required by charge conservation.

The corresponding outgoing
amplitudes are expressed as a linear combination of electron-like and hole-like waves as
\begin{equation}
\begin{pmatrix}
u_{S+} \\ v_{S+}
\end{pmatrix}
=C_3
\begin{pmatrix}
u_e(z) \\ v_e(z)
\end{pmatrix}, \quad
\begin{pmatrix}
u_{S-} \\ v_{S-}
\end{pmatrix}
=C_4
\begin{pmatrix}
u_h(z) \\ v_h(z)
\label{uS1}
\end{pmatrix},
\end{equation}
For the chosen real order parameter $\Delta$ the hole amplitudes are linked to the electron ones by means of the relations
\begin{equation}
u_h(z) = v^*_e(z), \quad v_h(z) = u^*_e(z)
\end{equation}
enabling one to express the wave functions \eqref{uS1} only in terms of the functions $u_e$ and $v_e$.
These functions can be found from the quasiclassical (Andreev) equation
\begin{equation}
\begin{pmatrix}
-iv_z\partial_z & \Delta \\
\Delta & iv_z\partial_z
\end{pmatrix}
\begin{pmatrix}
u_e \\ v_e
\end{pmatrix}
=\varepsilon
\begin{pmatrix}
u_e \\ v_e
\end{pmatrix},
\label{Andeq}
\end{equation}
combined with the asymptotic behavior deep in the superconducting bulk ($z\rightarrow\infty$)
\begin{equation}
\begin{pmatrix}
u_e(z) \\ v_e(z)
\end{pmatrix}
\sim
\begin{cases}
e^{i\sqrt{\varepsilon^2 - \Delta^2}z/v_z}, & \varepsilon > \Delta,
\\
e^{-\sqrt{\Delta^2 - \varepsilon^2}z/v_z}, & 0 < \varepsilon < \Delta.
\end{cases}
\end{equation}
As a result, one can derive the transmission probabilities for these two processes $\mathcal{D}^{e-e}_{NS,\sigma}$ and $\mathcal{D}^{e-h}_{NS,\sigma}$.

After a straightforward calculation (see appendix) we obtain
\begin{gather}
\mathcal{R}^{e-e}_{NS,\sigma}=
\left|u_e^2(0)\sqrt{R_{\sigma}} -
v_e^2(0)\sqrt{R_{-\sigma}}e^{i\sigma\theta}\right|^2
\mathcal{L}_\sigma,
\label{ReeNS}
\\
\mathcal{R}^{e-h}_{NS,\sigma}=
|u_e(0)|^2 |v_e(0)|^2 D_{\uparrow}D_{\downarrow}\mathcal{L}_\sigma,
\label{RehNS}
\\
\mathcal{D}^{e-e}_{NS,\sigma}=
\left[|u_e(0)|^2 - |v_e(0)|^2\right] |u_e(0)|^2 D_{\sigma}\mathcal{L}_\sigma,
\label{DeeNS}
\\
\mathcal{D}^{e-h}_{NS,\sigma}=
\left[|u_e(0)|^2 - |v_e(0)|^2\right] |v_e(0)|^2 R_{-\sigma}D_{\sigma} \mathcal{L}_\sigma,
\label{DehNS}
\end{gather}
where $\sigma =\pm$, $D_{\uparrow(\downarrow)}=1-R_{\uparrow(\downarrow)}$ is the normal state transmission probability for a spin-up (spin-down) electron,
\begin{equation}
\mathcal{L}_\sigma = \left|u_e^2(0) - v_e^2(0)
\sqrt{R_{\uparrow}R_{\downarrow}}e^{i\sigma\theta}\right|^{-2},
\label{Ls}
\end{equation}
and $u_e(0)$ and $v_e(0)$ are the interface values of the Bogolubov amplitudes. With the aid of the above expressions it is easy to
verify that the total scattering probability for an incident electron in Fig. 2a equals to one:
\begin{equation}
\mathcal{R}^{e-e}_{NS,\sigma} + \mathcal{R}^{e-h}_{NS,\sigma} +
\mathcal{D}^{e-e}_{NS,\sigma} + \mathcal{D}^{e-h}_{NS,\sigma}=1.
\end{equation}

The remaining 12 scattering processes in Fig. 2 can be treated analogously. For instance, the reflection and transmission probabilities
for the scattering processes of a hole-like excitation depicted in Fig. 2b read
\begin{gather}
\mathcal{R}^{h-h}_{NS,\sigma}=
\left|u_e^2(0)\sqrt{R_{-\sigma}} -
v_e^2(0)\sqrt{R_{\sigma}}e^{i\sigma\theta}\right|^2\mathcal{L}_\sigma ,
\label{RhhNS}
\\
\mathcal{R}^{h-e}_{NS,\sigma}=
 |u_e(0)|^2 |v_e(0)|^2 D_{\uparrow}D_{\downarrow}\mathcal{L}_\sigma,
\label{RheNS}
\\
\mathcal{D}^{h-h}_{NS,\sigma}=
\left[|u_e(0)|^2 - |v_e(0)|^2\right] |u_e(0)|^2
 D_{-\sigma}\mathcal{L}_\sigma,
\label{DhhNS}
\\
\mathcal{D}^{h-e}_{NS,\sigma}=
\left[|u_e(0)|^2 - |v_e(0)|^2\right] |v_e(0)|^2 R_{\sigma}D_{-\sigma}\mathcal{L}_\sigma ,
\label{DheNS}
\end{gather}
The scattering probabilities for electrons and holes coming from the 
superconductor (Fig. 2c and 2d) are specified in appendix.

Let us briefly analyze the above results. To begin with, we notice that in the case of spin-independent scattering $R_{\uparrow}=R_{\downarrow}$ and $\theta=0$
our Eqs. \eqref{ReeNS}-\eqref{DehNS} and \eqref{RhhNS}-\eqref{DheNS}
reduce to the standard BTK results \cite{Blonder82}. In this case both transmission and reflection probabilities remain symmetric under the
replacement of an electron by a hole and vice versa, i.e. we have, e.g., $\mathcal{R}^{e-e}_{NS,\sigma} = \mathcal{R}^{h-h}_{NS,\sigma}$,
$\mathcal{R}^{e-h}_{NS,\sigma} = \mathcal{R}^{h-e}_{NS,\sigma}$ and so on. These observations just confirm that no electron-hole asymmetry can be
induced by spin-independent scattering at the SN interface. Turning now to spin-sensitive scattering considered here we notice that scattering
probabilities are in general not anymore equal to each other. Comparing, for instance, Eqs. \eqref{ReeNS}-\eqref{DehNS} and \eqref{RhhNS}-\eqref{DheNS},
we observe that for  $R_{\uparrow} \neq R_{\downarrow}$ and $\theta \neq 0$ only two reflection probabilities remain equal,
$\mathcal{R}^{e-h}_{NS,\sigma}=\mathcal{R}^{h-e}_{NS,\sigma}$, whereas all others differ, e.g., $\mathcal{R}^{e-e}_{NS,+} \neq \mathcal{R}^{e-e}_{NS,-}$,
$\mathcal{R}^{e-e}_{NS,\sigma} \neq \mathcal{R}^{h-h}_{NS,\sigma}$, $\mathcal{R}^{e-h}_{NS,+}\neq\mathcal{R}^{h-e}_{NS,-}$, etc. Thus, we arrive
at an important conclusion: {\it Spin-sensitive quasiparticle scattering generates electron-hole imbalance in superconducting structures} which manifests itself
in different scattering rates for electrons and holes in such systems.

This conclusion has important implications for the thermoelectric effect in superconductors. As we already pointed out, electron-hole imbalance can be
considered as an important prerequisite for strong enhancement of the thermoelectric coefficient, see Eq.  \eqref{mott}. Below we will explicitly evaluate
thermoelectric currents in SN bilayers with a spin-active interfaces and demonstrate that an asymmetry in the scattering rates for electrons and holes
indeed yields large thermoelectric effect in such systems.

\section{Thermoelectric currents}
Making use of the above results for the 
quasiparticle wave functions and employing Eq. \eqref{tok} together with the 
normalization condition \eqref{norm} we can now evaluate
thermoelectric currents both in the superconductor ($z>0$) and in the normal metal ($z<0$). As these currents are directed along the SN interface,
below we will only be interested in the $x$-component of the current density $j_x(z)$.  Expressing the current in the superconductor in terms of both
reflection and transmission probabilities as well as quasiparticle distribution functions for the left and right movers at $x\rightarrow\pm\infty$, we obtain
\begin{multline}
j_x(z>0)=
-\dfrac{e}{2m}
\int\limits_0^{\infty}\dfrac{d\varepsilon}{2\pi}
\left[\tanh\dfrac{\varepsilon}{2 T_1} - \tanh\dfrac{\varepsilon}{2 T_2}\right]
\\
\times
\int\limits_{\substack{|\bm{p}_{\parallel}| < p_F \\ p_x>0}}
\dfrac{d^2 \bm{p}_{\parallel}}{(2\pi)^2}
\dfrac{p_x}{v_z}\dfrac{|u_e(z)|^2 + |v_e(z)|^2}{|u_e(0)|^2 - |v_e(0)|^2}
\\
\times
\sum_{\sigma =\pm}
\left(
\mathcal{R}_{SN,\sigma}^{e-e} + \mathcal{R}_{SN,\sigma}^{e-h}
- \mathcal{R}_{SN,\sigma}^{h-h} - \mathcal{R}_{SN,\sigma}^{h-e} \right.
\\
\left.
+ \mathcal{D}_{NS,\sigma}^{e-e} + \mathcal{D}_{NS,\sigma}^{e-h}
- \mathcal{D}_{NS,\sigma}^{h-h} - \mathcal{D}_{NS,\sigma}^{h-e}
\right).
\end{multline}
A similar expression can also be derived for the thermoelectric current in the normal metal. Combining both these expressions with
our results for the transmission and reflection probabilities, we finally get
\begin{multline}
\label{jsn}
j_x(z)=
\dfrac{e}{m}
\int\limits_0^{\infty}\dfrac{d\varepsilon}{2\pi}
\left[\tanh\dfrac{\varepsilon}{2 T_1} - \tanh\dfrac{\varepsilon}{2 T_2}\right]
\\
\times
\int\limits_{\substack{|\bm{p}_{\parallel}| < p_F \\ p_x>0}}
\dfrac{d^2 \bm{p}_{\parallel}}{(2\pi)^2}
\dfrac{p_x}{v_z}
|v_e(0)|^2 (R_{\uparrow} - R_{\downarrow})
(\mathcal{L}_{+} - \mathcal{L}_{-}) \mathcal{U}(z),
\end{multline}
where we defined
\begin{equation}
\mathcal{U}(z)=
\begin{cases}
|u_e(z)|^2 + |v_e(z)|^2, & z > 0,
\\
|v_e(0)|^2 - |u_e(0)|^2, & z < 0.
\end{cases}
\label{U}\end{equation}

Eqs. \eqref{jsn} and \eqref{U} represent the key result of this work. We observe that the thermoelectric current
vanishes identically \cite{FN} provided at least one of the two conditions, $R_{\uparrow}=R_{\downarrow}$ or $\theta=0$, is fulfilled.
If, however, both these conditions are violated, the thermoelectric current differs from zero and can become large.

Let us briefly analyze the above results. In the superconducting layer ($z>0$) the
thermoelectric current density \eqref{jsn}, \eqref{U} depends on the coordinate $z$ in the vicinity of the interface and
tends to some nonzero value in the bulk. In the normal metal, in contrast, $j_x$ remains spatially constant,
i.e. it does not depend on the distance $|z|$ from the interface. This a well known property of the ballistic model employed
here \cite{Z82}. Within this model the electron elastic mean free path $\ell$ tends to infinity and no electron momentum relaxation
occurs. Relaxing this condition, i.e. assuming the mean free path to be finite, one can demonstrate that $j_x(z)$ decays exponentially
into the normal metal at distances of order $\ell$. Hence, in this case the thermoelectric current is essentially confined to the
SN interface. The analysis of this physical situation is beyond the frames of this work and will be published elsewhere \cite{KZ14}.

In order to explicitly evaluate the thermoelectric current it is necessary to selfconsistently determine both
the functions $u_e(z)$, $v_e(z)$ and the order parameter $\Delta (z)$ for any given values of the parameters
$R_{\uparrow}$, $R_{\downarrow}$ and $\theta$. If, for simplicity, one
neglects the coordinate dependence of the order parameter by setting $\Delta(z>0)=\Delta$, one readily finds
\begin{equation}
\begin{pmatrix}
u_e(z) \\ v_e(z)
\end{pmatrix}
=
\begin{pmatrix}
\varepsilon + \sqrt{\varepsilon^2 - \Delta^2} \\ \Delta
\end{pmatrix}
e^{i\sqrt{\varepsilon^2 - \Delta^2}z/v_z},
\end{equation}
where we define $\Img \sqrt{\varepsilon^2 - \Delta^2} > 0$ for $\varepsilon^2 < \Delta^2$.
Combining these expressions with Eqs. \eqref{jsn}, \eqref{U} and splitting the energy integral
in Eq. \eqref{jsn} into subgap ($|\varepsilon | <\Delta$) and overgap ($|\varepsilon | >\Delta$)
parts, one observes that the overgap contribution to the current vanishes because the condition
$\mathcal{L}_+ = \mathcal{L}_-$ holds under this approximation. The subgap contribution to $j_x$
also vanishes in the normal metal and remains non-zero in the superconductor in the vicinity of
the SN interface.

The subgap contribution to $j_x$ shows the same behavior also if one relaxes the condition
$\Delta(z>0)=\Delta$ and takes into account the proximity induced suppression of the order
parameter $\Delta (z)$ near the SN interface. In this case $\mathcal{L}_+$ does not in general
coincide with $\mathcal{L}_-$ and, hence, the overgap contribution to the thermoelectic current
differs from zero both in normal and superconducting layers.

Estimating the magnitude of the thermoelectric current density at intermediate temperatures $T_1,T_2\sim\Delta$,
from Eqs. \eqref{jsn} and \eqref{U} we obtain
\begin{equation}
j_x\sim e v_F N_0 (R_{\uparrow} - R_{\downarrow}) \sin\theta (T_1 - T_2),
\label{jT}
\end{equation}
where $N_0\equiv N(\varepsilon_F)=m p_F/(2\pi^2)$ is the normal state density of states at the Fermi level.
In contrast to the standard result \cite{Galperin73}, the expression \eqref{jT}
does not contain the small factor $T/\varepsilon_F \ll 1$, i.e. the thermoelectric effect can be large
If one furthermore sets $(R_{\uparrow} - R_{\downarrow}) \sin\theta \sim 1$ and $T_1 - T_2 \sim T_c$, the thermoelectric
current density \eqref{jT} becomes of the same order as the critical one for a clean superconductor $j_x \sim j_c \sim  e v_F N_0 T_c$.

In summary, we demonstrated that quasiparticle scattering at spin-active interfaces is characterized by different scattering rates
for electrons and holes, thus being responsible for electron-hole imbalance generation in superconducting hybrids under consideration.
As a result of this imbalance, the thermoelectric currents in such structures can be greatly enhanced and under certain conditions
may reach remarkably high values of order of the critical (depairing) current of a superconductor. This thermoelectric effect
can be reliably detected in modern experiments with bimetallic superconducting 
rings (see, e.g., Refs. \onlinecite{Zavaritskii74,Falco76,Harlingen80,Pe}
and a discussion in Ref. \onlinecite{Kalenkov12}) and can be exploited in a 
number of novel devices, such as, e.g., thermoelectric bolometers.

\appendix

\section{Bogolyubov wave functions}

Resolving Bogolubov equations \eqref{BdG} with appropriate boundary and 
asymptotic conditions, we derive explicit expressions for the quasiparticle and 
hole wave functions in the S- and N-parts of our bilayer. In general, the 
Bogolubov amplitudes $u$, $v$ have the form of the following two component 
spinors
\begin{equation}
\begin{pmatrix}
u \\ v
\end{pmatrix}
=
\begin{pmatrix}
u_{\uparrow}  \\ u_{\downarrow} \\ v_{\uparrow}  \\ v_{\downarrow}
\end{pmatrix}.
\end{equation}
Within our model, interface electron scattering preserves its spin
projection. Hence, the solutions of the Bogolubov equations
can be split into two different classes,
\begin{equation}
\begin{pmatrix}
u \\ v
\end{pmatrix}
=
\begin{pmatrix}
u_{\uparrow}  \\ 0 \\ v_{\uparrow}  \\ 0
\end{pmatrix}
\text{\ and }
\begin{pmatrix}
u \\ v
\end{pmatrix}
=
\begin{pmatrix}
0  \\ u_{\downarrow} \\ 0  \\ v_{\downarrow}
\end{pmatrix},
\end{equation}
describing respectively spin-up and spin-down excitations in our structure. For
the sake of simplicity, here we will indicate only nonzero components of the
corresponding Bogolubov amplitudes.

As illustrated in Fig. 2, all scattering processes can be classified in four
different groups (a), (b), (c) and (d) depending on whether incident
electron-like or hole-like excitations come from the normal metal or
superconductor. For each of these four groups
one can evaluate the corresponding wave functions and obtain:
\begin{widetext}

(a) The wave function describing scattering of an electron-like excitation 
coming from the bulk of the normal metal reads
\begin{gather}
S:\quad
\begin{pmatrix}
u_e(z) \\ v_e(z)
\end{pmatrix}
\dfrac{i\sqrt{D_{\sigma}}u_e(0)e^{i\sigma\theta /2}}{
u_e^2(0) - v_e^2(0) \sqrt{R_{\uparrow}R_{\downarrow}}e^{i\sigma\sigma}}
e^{ip_{z}z}e^{i\bm{p}_{\parallel}\bm{\rho}}
+
\begin{pmatrix}
v_e(z) \\ u_e(z)
\end{pmatrix}
\dfrac{i\sqrt{R_{-\sigma}D_{\sigma}}v_e(0) e^{i\sigma\theta}}{
u_e^2(0) - v_e^2(0) \sqrt{R_{\uparrow}R_{\downarrow}}e^{i\sigma\theta}}
e^{-ip_{z}z}e^{i\bm{p}_{\parallel}\bm{\rho}},
\\
N:\quad
\begin{pmatrix}
1 \\0
\end{pmatrix}
e^{i\varepsilon z / |v_{z}|} e^{ip_{z}z} e^{i\bm{p}_{\parallel}\bm{\rho}}
+
\begin{pmatrix}
e^{i\sigma\theta/2}\dfrac{
\sqrt{R_{\sigma}} u_e^2(0) - \sqrt{R_{-\sigma}}v_e^2(0)e^{i\sigma\theta}
}{u_e^2(0) - v_e^2(0) \sqrt{R_{\uparrow}R_{\downarrow}}e^{i\sigma\theta}}
e^{-i\varepsilon z / |v_{z}|} e^{-ip_{z}z}
\\
\dfrac{\sqrt{D_{\uparrow}D_{\downarrow}} u_e(0) v_e(0)e^{i\sigma\theta}
}{u_e^2(0) - v_e^2(0) \sqrt{R_{\uparrow}R_{\downarrow}}e^{i\sigma\theta}}
e^{-i\varepsilon z / |v_{z}|} e^{ip_{z}z}
\end{pmatrix}e^{i\bm{p}_{\parallel}\bm{\rho}},
\end{gather}
(b) For the wave function describing scattering of a hole-like excitation coming 
from the bulk of the normal metal we obtain
\begin{gather}
S:\quad
-
\begin{pmatrix}
u_e(z) \\ v_e(z)
\end{pmatrix}
\dfrac{i\sqrt{R_{\sigma}D_{-\sigma}}v_e(0) e^{i\sigma\theta}}{
u_e^2(0) - v_e^2(0) \sqrt{R_{\uparrow}R_{\downarrow}}e^{i\sigma\theta}}
e^{ip_{z}z} e^{i\bm{p}_{\parallel}\bm{\rho}}
-
\begin{pmatrix}
v_e(z) \\ u_e(z)
\end{pmatrix}
\dfrac{i\sqrt{D_{-\sigma}}u_e(0)e^{i\theta_{\sigma}/2}}{u_e^2(0) - v_e^2(0) 
\sqrt{R_{\uparrow}R_{\downarrow}}e^{i\sigma\theta}}
e^{-ip_{z}z} e^{i\bm{p}_{\parallel}\bm{\rho}},
\\
N:\quad
\begin{pmatrix}
0 \\1
\end{pmatrix}
e^{i\varepsilon z / |v_{z}|} e^{-ip_{z}z} e^{i\bm{p}_{\parallel}\bm{\rho}}
+
\begin{pmatrix}
\dfrac{\sqrt{D_{\uparrow}D_{\downarrow}} u_e(0) v_e(0)e^{i\sigma\theta}
}{u_e^2(0) - v_e^2(0) \sqrt{R_{\uparrow}R_{\downarrow}}e^{i\sigma\theta}}
e^{-i\varepsilon z / |v_{z}|} e^{-ip_{z}z}
\\
e^{i\sigma\theta/2}\dfrac{\sqrt{R_{-\sigma}} u_e^2(0) - 
\sqrt{R_{\sigma}}v_e^2(0)e^{i\sigma\theta}
}{u_e^2(0) - v_e^2(0) \sqrt{R_{\uparrow}R_{\downarrow}}e^{i\sigma\theta}}
e^{-i\varepsilon z / |v_{z}|} e^{ip_{z}z}
\end{pmatrix} e^{i\bm{p}_{\parallel}\bm{\rho}},
\end{gather}
(c) The wave function describing scattering of an electron-like excitation 
coming from the superconductor bulk has the form
\begin{multline}
S:\quad
\begin{pmatrix}
v_h(z) \\ u_h(z)
\end{pmatrix}
e^{-ip_{z}z} e^{i\bm{p}_{\parallel}\bm{\rho}}
+
\begin{pmatrix}
u_e(z) \\ v_e(z)
\end{pmatrix}
\dfrac{u_e(0)v_h(0) - u_h(0)v_e(0)}{u_e^2(0) - v_e^2(0) 
\sqrt{R_{\uparrow}R_{\downarrow}}e^{i\sigma\theta}}
\sqrt{R_{\sigma}} e^{i\sigma\theta/2}
e^{ip_{z}z} e^{i\bm{p}_{\parallel}\bm{\rho}}
-\\-
\begin{pmatrix}
v_e(z) \\ u_e(z)
\end{pmatrix}
\dfrac{u_e(0) u_h(0) - v_e(0) v_h(0) \sqrt{R_{\uparrow}R_{\downarrow}}e^{i/2}
e^{i\sigma\theta}}{
u_e^2(0) - v_e^2(0) \sqrt{R_{\uparrow}R_{\downarrow}}e^{i\sigma\theta}}
e^{-ip_{z}z} e^{i\bm{p}_{\parallel}\bm{\rho}},
\end{multline}
\begin{equation}
N:\quad
\dfrac{u_e(0)v_h(0) - u_h(0)v_e(0)}{u_e^2(0) - v_e^2(0) 
\sqrt{R_{\uparrow}R_{\downarrow}}e^{i\sigma\theta}}
\begin{pmatrix}
i\sqrt{D_{\sigma}} u_e(0) e^{i\sigma\theta/2} e^{-i\varepsilon z / |v_{z}|} 
e^{-ip_{z}z}
\\
i\sqrt{R_{\sigma}D_{-\sigma}} v_e(0) e^{i\sigma\theta} e^{-i\varepsilon z / 
|v_{z}|} e^{ip_{z}z}
\end{pmatrix} e^{i\bm{p}_{\parallel}\bm{\rho}}.
\end{equation}
(d) For the wave function describing scattering of a hole-like excitation coming 
from the superconducting bulk we find
\begin{multline}
S:\quad
\begin{pmatrix}
u_h(z) \\ v_h(z)
\end{pmatrix}
e^{ip_{z}z} e^{i\bm{p}_{\parallel}\bm{\rho}}
-
\begin{pmatrix}
u_e(z) \\ v_e(z)
\end{pmatrix}
\dfrac{u_e(0) u_h(0) - v_e(0) v_h(0) 
\sqrt{R_{\uparrow}R_{\downarrow}}e^{i\sigma\theta}}{
u_e^2(0) - v_e^2(0) \sqrt{R_{\uparrow}R_{\downarrow}}e^{i\sigma\theta}}
e^{ip_{z}z} e^{i\bm{p}_{\parallel}\bm{\rho}},
-\\-
\begin{pmatrix}
v_e(z) \\ u_e(z)
\end{pmatrix}
\dfrac{u_e(0)v_h(0) - u_h(0)v_e(0)}{u_e^2(0) - v_e^2(0) 
\sqrt{R_{\uparrow}R_{\downarrow}}e^{i\sigma\theta}}
\sqrt{R_{-\sigma}} e^{i\sigma\theta/2}
e^{-ip_{z}z} e^{i\bm{p}_{\parallel}\bm{\rho}}
\end{multline}
\begin{equation}
N:\quad
\dfrac{u_e(0)v_h(0) - u_h(0)v_e(0)}{u_e^2(0) - v_e^2(0) 
\sqrt{R_{\uparrow}R_{\downarrow}}e^{i\sigma\theta}}
\begin{pmatrix}
i\sqrt{R_{-\sigma}D_{\sigma}} v_e(0) e^{i\sigma\theta} e^{-i\varepsilon z / 
|v_{z}|} e^{-ip_{z}z}
\\
-i\sqrt{D_{-\sigma}} u_e(0) e^{i\sigma\theta/2} e^{-i\varepsilon z / |v_{z}|} 
e^{ip_{z}z}
\end{pmatrix} e^{i\bm{p}_{\parallel}\bm{\rho}}.
\end{equation}
\end{widetext}
Index $\sigma$ distinguishes spin-up and spin-down wave functions.

Making use of the above expressions we recover both normal and Andreev 
reflection and transmission probabilities for all 16 processes depicted
in Fig. 2. Eqs. \eqref{ReeNS}-\eqref{DehNS} and 
\eqref{RhhNS}-\eqref{DheNS} define scattering probabilities for 8 of these 
processes.
The remaining 8 probabilities are:
\begin{gather}
\mathcal{R}^{e-e}_{SN,\sigma}=
\left[|u_e(0)|^2 - |v_e(0)|^2\right]^2
R_{\sigma}\mathcal{L}_{\sigma},
\\
\mathcal{R}^{e-h}_{SN,\sigma}=
\left|u_e(0) v_e^*(0) - v_e(0) u_e^*(0)
\sqrt{R_{\uparrow}R_{\downarrow}}e^{i\sigma\theta}\right|^2\mathcal{L}_{\sigma},
\\
\mathcal{D}^{e-e}_{SN,\sigma}=
\left[|u_e(0)|^2 - |v_e(0)|^2\right] |u_e(0)|^2
D_{\sigma}\mathcal{L}_{\sigma} ,
\\
\mathcal{D}^{e-h}_{SN,\sigma}=
\left[|u_e(0)|^2 - |v_e(0)|^2\right] |v_e(0)|^2
R_{\sigma}D_{-\sigma} \mathcal{L}_{\sigma},
\end{gather}
\begin{gather}
\mathcal{R}^{h-h}_{SN,\sigma}=
\left[|u_e(0)|^2 - |v_e(0)|^2\right]^2
R_{-\sigma}\mathcal{L}_{\sigma},
\\
\mathcal{R}^{h-e}_{SN,\sigma}=
\left|u_e(0) v_e^*(0) - v_e(0) u_e^*(0)
\sqrt{R_{\uparrow}R_{\downarrow}}e^{i\sigma\theta}\right|^2
\mathcal{L}_{\sigma},
\\
\mathcal{D}^{h-h}_{SN,\sigma}=
\left[|u_e(0)|^2 - |v_e(0)|^2\right] |u_e(0)|^2
D_{-\sigma} \mathcal{L}_{\sigma},
\\
\mathcal{D}^{h-e}_{SN,\sigma}=
\left[|u_e(0)|^2 - |v_e(0)|^2\right] |v_e(0)|^2
R_{-\sigma}D_{\sigma} \mathcal{L}_{\sigma},
\end{gather}
where $\mathcal{L}_{\sigma}$ is again defined in Eq. \eqref{Ls}.

In order to evaluate the electric current in our system it is necessary to
properly normalize the above wave functions. This task can be accomplished with 
the aid of Eq. \eqref{norm}. The wave functions describing scattering of 
electron-like and hole-like excitations coming from the superconductor bulk obey 
the following normalization condition
\begin{multline}
\int
\left[
u^*_{\bm{p}_{\parallel},\varepsilon}(\bm{r})
u_{\bm{p}_{\parallel}^{\prime},\varepsilon'}(\bm{r})
+
v^*_{\bm{p}_{\parallel},\varepsilon}(\bm{r})
v_{\bm{p}_{\parallel}^{\prime},\varepsilon'}(\bm{r})
\right]
d\bm{r}
=(2\pi)^3
\\\times
|v_x|
\frac{\sqrt{\varepsilon^2 - \Delta^2}}{\varepsilon} 
\delta(\varepsilon - \varepsilon')
\delta(\bm{p}_{\parallel} - \bm{p}_{\parallel}^{\prime}).
\end{multline}
At the same time, the normalization condition for the wave functions of  
electrons and holes coming from the side of the normal metal take a slightly 
different form, i.e.
\begin{multline}
\int
\left[
u^*_{\bm{p}_{\parallel},\varepsilon}(\bm{r})
u_{\bm{p}_{\parallel}^{\prime},\varepsilon'}(\bm{r})
+
v^*_{\bm{p}_{\parallel},\varepsilon}(\bm{r})
v_{\bm{p}_{\parallel}^{\prime},\varepsilon'}(\bm{r})
\right]
d\bm{r}
\\=(2\pi)^3 |v_x|
\delta(\varepsilon - \varepsilon')
\delta(\bm{p}_{\parallel} - \bm{p}_{\parallel}^{\prime}).
\end{multline}


\begin{thebibliography}{99}
\bibitem{FN} Here we neglect a small contribution to the thermoelectric coefficient \cite{Galperin73}.
\end{thebibliography}
\end{document}